\documentclass[RNAAS]{aastex62}

\graphicspath{}

\received{July 23, 2018}
\revised{July 23, 2018}
\accepted{\today}
\submitjournal{RNAAS}

\shorttitle{}
\shortauthors{Gordon, Pimbblet \& Owers.}

\begin{document}

\title{The Impact of Spectroscopic Fibre Collisions on the Observed Angular Correlation Function}

\correspondingauthor{Yjan A. Gordon}
\email{yjan.gordon@umanitoba.ca}

\author[0000-0003-1432-253X]{Yjan A. Gordon}
\affiliation{Department of Physics and Astronomy,
University of Manitoba,
Winnipeg, MB,
R3T 2N2, Canada}
\affiliation{E.A. Milne Centre for Astrophysics,
University of Hull,
Cottingham Road,
Kingston-upon-Hull,
HU6 7RX, UK}

\author{Kevin A. Pimbblet}
\affiliation{E.A. Milne Centre for Astrophysics,
University of Hull,
Cottingham Road,
Kingston-upon-Hull,
HU6 7RX, UK}

\author{Matt S. Owers}
\affiliation{Department of Physics and Astronomy,
Macquarie University,
NSW 2109, Australia}
\affiliation{Australian Astronomical Observatory,
105 Dehli Road, North Ryde,
NSW 2113, Australia}



\section{}

One of the complications associated with large-scale spectroscopic galaxy surveys is that of fibre collisions \citep{Strauss2002, Patton2008}.
That is to say, the physical size of a spectroscopic fibre plug prevents observations of targets that are too close to each other on the sky.
These fibre collisions may thus hinder studies of regions with a high target density, such as interacting galaxy pairs or galaxy clusters \citep[e.g.,][]{Robotham2010,
Gordon2017}.
Comparisons with mock catalogues can allow for the recovery of clustering lost to fibre collisions by the application of statistical corrections (e.g., \citealp{Guo2012} and \citealp{Hahn2017}).
Alternatively, the observing strategy of some surveys is designed to circumvent the negative impact of fibre collisions by re-observing the same field using different fibre configurations \citep{Driver2011}.
In order to quantify the effect of this latter technique, we compare the 2-point angular correlation function (2PCF) of a survey that adopts this strategy to one which does not.

The spectroscopic component of the Sloan Digital Sky Survey \citep[SDSS,][]{York2000} is a single-pass survey with a $55''$ fibre collision limit, and $94\,$\% completeness down to $r_{\rm{Petrosian}}<17.77$ \citep{Strauss2002}.
This completeness level is dependent on the target density however, with some observed clusters in SDSS having just $65\,$\% spectroscopic completeness \citep{Yoon2008}.
Conversely, the Galaxy And Mass Assembly survey \citep[GAMA,][]{Driver2011} uses a multi-pass strategy for its spectroscopic campaign, achieving $98.5\,$\% completeness at $r_{\rm{Petrosian}}<19.8$ that is broadly independent of the target density \citep{Liske2015}.
Our analysis uses galaxies observed up to data release 10 of SDSS \citep{Ahn2014} and data release 3 of GAMA \citep{Baldry2018} residing within the GAMA equatorial fields G12 and G15.
These fields lie entirely within the GAMA and SDSS survey footprints, allowing for the analysis to be conducted on the same region of sky.
To prevent differing survey depths biasing our result, only galaxies with $r_{\rm{Petrosian}} < 17.77$ are included.

\begin{figure}[]
\centering
\includegraphics[width=0.9\columnwidth]{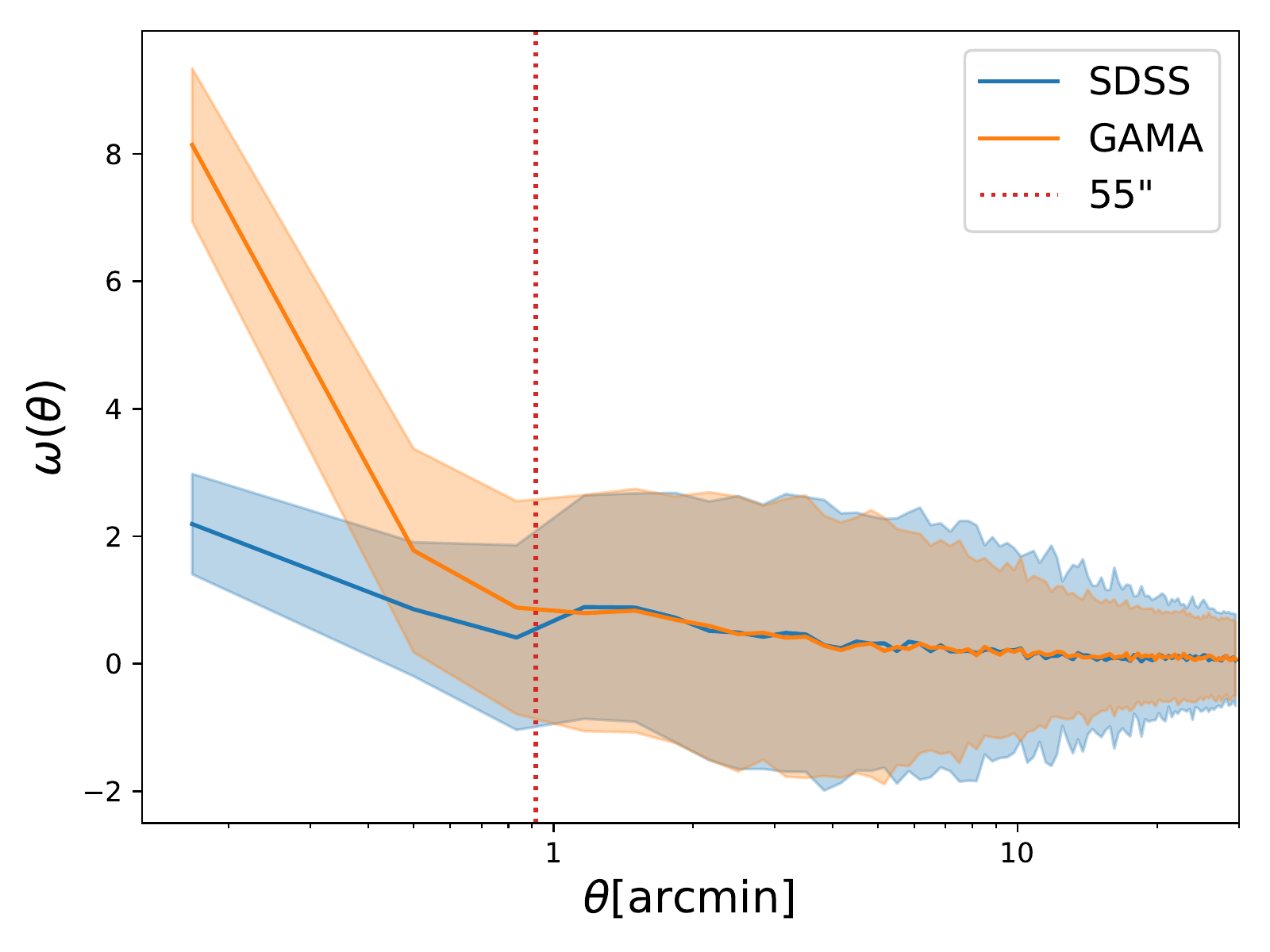}
\caption{The 2-point angular correlation functions, $\omega(\theta)$, of the SDSS (blue) and GAMA (orange) spectroscopic surveys over the GAMA fields G12 and G15. A Landy-Szalay estimator was used and the errors are estimated using the bootstrap method. The red dashed line shows the SDSS fibre collision limit of $55''$. There is a clear ($>4\sigma$ confidence) increase in GAMA detections on the smallest angular scales.}
\label{2pcf}
\end{figure}

To determine the angular correlation function we produce mock galaxy populations equal in size to the GAMA and SDSS populations, and distribute the mock galaxies randomly along the same spatial area occupied by the real observations.
Each galaxy is then paired with all other galaxies observed by the same survey inside a $0.5^o$ radius.
The Landy-Szalay estimator, $\omega(\theta)$, given by
\begin{equation}
\omega(\theta) = \frac{DD - 2DR + RR}{RR},
\end{equation}
where $DD$, $DR$, and $RR$, are the data-data, data-random, and random-random pairs respectively, is then used to calculate the angular correlation \citep{Landy1993}. 
In Figure \ref{2pcf}, we show the 2PCFs and their uncertainties, estimated by the bootstrap method, for both GAMA and SDSS in pair separation bins of $20''$ \citep[$\sim40\,$kpc at the median SDSS redshift of $z=0.1$,][]{Abazajian2009} out to to $0.5^o$.
The $55''$ SDSS fibre collision limit is shown for reference. The angular correlation functions are shown to deviate at separations below the SDSS fibre collision limit, reaching a $> 4\sigma$ significance at pair separations of $<20''$.
This unambiguously demonstrates the merit of field re-observation in spectroscopic surveys where studying regions of high galaxy density is a priority.

\acknowledgements
YAG acknowledges funding from a University of Hull PhD studentship.
KAP acknowledges the support of Science and Technology Facilities Council (STFC), through the University of Hull's Consolidated Grant ST/R000840/1.
MSO acknowledges the funding support from the Australian Research Council through a Future Fellowship (FT140100255)

GAMA is a joint European-Australasian project based around a spectroscopic campaign using the Anglo-Australian Telescope. The GAMA input catalogue is based on data taken from the Sloan Digital Sky Survey and the UKIRT Infrared Deep Sky Survey. Complementary imaging of the GAMA regions is being obtained by a number of independent survey programmes including GALEX MIS, VST KiDS, VISTA VIKING, WISE, Herschel-ATLAS, GMRT and ASKAP providing UV to radio coverage. GAMA is funded by the STFC (UK), the ARC (Australia), the AAO, and the participating institutions. The GAMA website is http://www.gama-survey.org/ .

\bibliographystyle{aasjournal}
\bibliography{$HOME/Documents/PHD_Work/Papers/bibTeX/library} 



\end{document}